# Mobile Security for the modern CEO: Attacks, Mitigations, and Future Trends

*Marc Schmitt* [a, 1]

[a] *Department of Computer Science, University of Oxford, UK*

**Abstract**

Today's world is digital, global, and interconnected and mobile devices are at the heart of modern communications in business, politics, and civil society. However, cyber threats are an omnipresent reality in our hyper-connected world. The world economic forum ranks cyber threats consistently among the global top security risks. Attacks on mobile devices grow yearly in volume and magnitude causing severe damage. This paper offers a comprehensive overview of modern mobile attacks categorized into malware, phishing, communication, supply chain, physical, and authentication attacks, including a section on mitigations and limitations. It also provides security design tips to secure the mobile setup and general recommendations to prevent the successful execution of an incoming attack. The last section highlights future technology trends and how those will impact and change the mobile security landscape in the future.

**Keywords:** Mobile system security; cyber security; cyber threats; malware; social engineering; CEO

[1] Corresponding author. E-mail address: marcschmitt@hotmail.de

# 1 Introduction

In the last decade, there has been an astonishing increase in connectivity, stored data, and advanced analytics. This digital storm has brought us another change, which is increasingly shaping the world – the smartphone. Several factors helped mobile devices to spread further. The most important ones are information democratization; the need or expectation for individuals and corporations to access that information in real-time; a significant shift from local markets towards a global economy; and a mindset shift from as soon as possible to right now (Weichbroth & Łysik, 2020).

Mobile devices have become an omnipresent reality and the importance of those devices will continue to increase (Weichbroth & Łysik, 2020). In addition, the external economic shock experienced in 2020 – the coronavirus pandemic – further accelerated the trend toward digitalization and the amount of mobile device utilization to access corporate systems. According to several studies, this trend will with a high likelihood gain momentum over the years (Zakaria, 2020). At the same time, cyber-attacks have been growing significantly and accelerated further (Delcher, 2021). The attack surface grew accordingly and made mobile devices more susceptible than ever to cyber threats (Check Point, 2021). Mobile devices have introduced a new and powerful gateway to the enterprise backend, and executives and other business stakeholders must recognize this new threat, which increases in importance due to the above-mentioned developments yearly. There is no "silver bullet" to stop cybercrime (Richardson, 2020), but giving up and losing the fight against cyber attackers is no option either. Organizations and individuals can shield themselves better from cyber-attacks by acquiring an understanding of the potential attack surface and possible mitigation techniques. This requires carrying out an objective assessment of security weaknesses and vulnerabilities including an identification of proper defenses.

In this paper, we analyze the mobile attack surface including mitigations based on the hypothetical scenario of a Chief Executive Officer (CEO) who is an active smartphone user. The goal is a comprehensive analysis, description, and discussion of the threats, attacks, and mitigation techniques including possible limitations. Most employees in today's world do not differentiate between smartphones for private and business purposes. Therefore – given our world of disappearing boundaries – we will consider attacks on the company as well as private information on the same device.

This paper is structured as follows: Section 2 "Mobile Threat Landscape" will define the mobile threat landscape and the scope of the analysis. Section 3 "Attacks and Mitigation" builds the major body of this paper. Here we will explain and discuss the different attacks/techniques that are utilized to achieve an adversary's goal, including specific mitigations and where relevant a discussion on the limitations. Section 4 "Security Design and Future Research" builds on the earlier analysis and gives recommendations for a modern mobile security design and also offers a best-practice summary for the CEO. In addition, hot topics in mobile security are introduced, including suggestions for future research. Section 5 ends with a summary and conclusion.

## 2   Mobile Threat Landscape

Our CEO is a high-profile target, which results in many potential adversaries that might be interested in gaining access to her smartphone. Also, mobile devices can serve as an entrance point for the enterprise backend and more serious security breaches/compromises thereafter. Once an adversary has access to the mobile device, he could initiate a lateral move into the enterprise and carry out additional attacks. A mobile device as our smartphone interacts via different protocols/communication networks with the internet and subsequently accesses public app stores, vendor infrastructures, social networks, websites, etc. The need to connect to (potentially unsecured) internet access points make mobile devices subject to interception attacks. Another option is the direct physical connection via a USB cable to another device, which could lead to physical security breaches. To give the analysis a more structured nature we will continue the discussion based on the following six core threat categories:

1. **Malware:** Malicious applications are one of the standard methods used by adversaries to gain control over mobile devices and can have a variety of different purposes. Malware is delivered via other methods/attack vectors as phishing, network attacks, or supply-chain compromises.

2. **Phishing:** This section covers social engineering attacks with a particular focus on spam phishing, spear-fishing, and smishing. Humans tend to be the weakest link in the security chain and are often an easy target.

3. **Communication:** This section covers all network-based attacks that impact communication channels (Wi-Fi, Bluetooth, and Cellular).

4. **Supply Chain:** This section covers physical supply chains as well as the mobile ecosystem – mainly the utilization of app stores to deliver malware.

5. **Physical and Authentication:** The last section covers physical device breaches caused by loss, theft, malicious charging stations, and other external devices including all authentication methods as passwords, pins, patterns, and biometrics to gain access to the mobile device.

The table below is based on the MITRE ATT&CK[2] list of attacks and matches all possibilities to associated sections in this paper.

**Figure 1 - Mapping of MITRE attack techniques to section**

| Category | Techniques | Covered in section |
| --- | --- | --- |
| Device Access | Deliver Malicious App via Authorized App Store | Malware, Supply Chain |
| | Deliver Malicious App via Other Means | Malware, Phishing |
| | Drive-by Compromise | Malware |
| | Exploit via Charging Station or PC | Physical and Authentication |
| | Exploit via Radio Interfaces | Communication |
| | Install Insecure or Malicious Configuration | Malware |
| | Lockscreen Bypass | Physical and Authentication |
| | Masquerade as Legitimate Application | Malware |
| | Supply Chain Compromise | Supply Chain |
| Network Attacks | Downgrade to Insecure Protocols | Communication |
| | Eavesdrop on Insecure Network Communication | Communication |
| | Exploit SS7 to Redirect Phone Calls/SMS | Communication |
| | Exploit SS7 to Track Device Location | Communication |
| | Jamming or Denial of Service | Communication |
| | Manipulate Device Communication | Communication |
| | Rogue Cellular Base Station | Communication |
| | Rogue Wi-Fi Access Points | Communication |

There are two big differences when it comes to the attack surface of mobile devices. The first obvious one is getting device access, which is probably one of the most important steps for most cyber-attacks. (MITRE, 2021) lists the usual attack options for gaining initial device access. The second one is to compromise the network or communication channels our smartphone is connected to. We can differentiate between threats or attacks that need and strive for device access and attacks that target exclusively the network and communication channels to cause harm without access to our device.

---

[2] MITRE ATT&CK® is a globally accessible knowledge base of adversary tactics and techniques based on real-world observations. ATT&CK is open and available to any person or organization for use at no charge.

The following section will focus on each of these categories and describe the associated attacks/techniques in more detail. Also, mitigation techniques will be discussed including their limitations.

## 3 Attacks and Mitigations

This part will focus on the defined threat categories and describe the different attacks for each. This includes examples, mitigation techniques, as well as potential limitations. Due to the interconnected nature of cybersecurity topics can overlap in some areas.

### 3.1 Malware

Due to the overlapping nature of security topics, we will start with an overview of malware types as malware is usually utilized during several concrete attack scenarios described later below. We can think of this chapter as the X we would move out of the brackets in an equation. Cyber-adversaries use malware to perform malicious activities on their behalf. Malware is designed to infiltrate, disrupt and/or damage computers without the user's consent. It can be present in any layer of the system stack – either as an independent program or embedded in another application or document (Lee, 2019).

#### 3.1.1 Malware Types

The term malware includes viruses, worms, Trojans, rootkits, spyware, adware, keyloggers, botnets, and more. Below is a taxonomy for the different malware types including a short description (Abijah Roseline & Geetha, 2021; Kaspersky, 2021b; Lee, 2019).

**Figure 2** – Malware Types

| Type | Description |
|---|---|
| Virus | Hidden, self-replicating software that propagates by infecting another program (parasitic infectors). A virus cannot replicate itself without the activation of a host/user. |
| Worm | Stand-alone malicious programs. Able to self-replicate and propagate independently. No user activation required to execute or spread their code. Able to make multiple copies of itself and spread across the network/internet. |
| Trojan | A non-replicating program that is disguised as a useful application, but performs a different function in the background without the knowledge of the user. E.g. Cryptojacking - mobile miners that utilize a device's processing power to mine cryptocurrencies or adware that displays additional advertisement. |
| Rootkit | Stealth software, which is actively flying under the radar. This is achieved by gaining privileges as root access. It then gains control over the devices functions (e.g. security software) |
| Backdoor | Allows unauthorized access. A back door is an alternative method of access to programs or hardware systems that bypasses the usual security mechanisms. |
| Spyware | Secretly installed software that invades the user's privacy by gathering information on individuals or organizations without their knowledge (e.g. Keylogger) |
| Adware | Software that shows the user advertising in addition to the actual function or installs additional software that displays advertising. |
| Ransomware | Also known as cryptoviral extortion. Ransomware gains access to the victims data and encrypts their files (private person or organization). The attacker then demands a ransom payment to deliver the encryption key. |
| Botnet | Infects a system and becomes a node. Many of thoes nodes build a group of automated malicious programs called bots. The botnet can be used to carry out differnet attacks (e.g. DDoS, distribution of viruses, and worms) |

Malware types that are especially relevant for mobile devices as Adware, Ransomware, Rootkits, and Trojans. Advertisement-related malware embedded in applications is one of the most widespread and also profitable enterprises today. It is not necessary very harmful to the end-user, but tends to be annoying and might drain the energy of the mobile device faster. Another modern malware type is crypto-jacking, which refers to unauthorized cryptocurrency mining on a victim's mobile device. This is done by embedding crypto-minder code in the app. The crypto-miners then utilize a device's processing power to mine cryptocurrencies (Bitcoin, Ethereum, or another altcoin). While not very harmful to the device owner it will constantly drain the device's energy as the crypto-miner will be active and independent from the host application (Abijah Roseline & Geetha, 2021).

### 3.1.2   Malware Distribution Channels

Malware needs to be delivered from the adversary to the target. Adversaries use different methods to install Malware or malicious configuration settings on our devices. There are several different ways to accomplish this objective. The simplest way is where the adversary gains access to a system through a user visiting a website over the normal course of browsing, which is rather random. Another way is to deliver malicious apps via **official app stores**. This is one of the best ways to do it as most mobile devices tend to be configured in a way that only allows installations from authorized app stores like the Google Play Store or the Apple App Store. If the attacker manages to place malware-infected applications onto those official stores it will be possible to distribute them easily. To do that, adversaries can use fake identities to create developer accounts to publish malicious applications to app stores. Also, once live on the app store the adversary can gain access to the target's Google account and use the possibility of remote installation to install this malicious app on an android device (MITRE, 2021). The obvious alternative to authorized app stores is are **3rd party app stores**. The advantage would be unlikely detection and easy setup due to less scrutiny compared to official app stores. The major downside of this approach is the fact that many mobile devices do not allow the installation of applications outside of the authorized app store ecosystem. A third option is to utilize social engineering as **phishing** emails or text messages to convince the target to visit the website. Phishing emails or text messages, either directly contain an attachment or use a weblink to finalize the attack (MITRE, 2021). More details on phishing can be found in section 3.2.

Before malware is delivered it needs to be packaged in a way to conceive its victims. Independent of the specific malware or distribution type, but especially for app-store delivery it makes sense to masquerade as a legitimate application. The closer the malicious application looks to the real one the easier it will be to convince the target to download and keep the malicious application on their devices. This is done by embedding the malware in a legitimate application or by closely aligning the design of the malicious app with a popular real app. Embedding the malware requires downloading and reengineering the application. Once this is done, the attacker can upload the malware-infected app to 3rd party or authorized app stores, or use other methods as phishing to deliver the infected app (MITRE, 2021).

The following two examples will show the danger of today's sophisticated malware packages. **Anubis** started as a tool for cyber espionage but was revised and entered the market again as a banking trojan. It is a fairly comprehensive malware package and can completely take over an Android mobile device. Anubis incorporates many different malware types as trojan, spyware, and even ransomware. One way Anubis was distributed was via phishing emails, that contained a link to a downloadable Android Package Kit (APK). This is the normal way to install Android applications. During the installation, the user was asked to Google Play Protect, which in reality did the opposite. This action disabled the real Google Play Protect and gave Anubis various permissions Anubis (Feller, 2020). See the table below.

**Figure 3 – Anubis Permission Rights**

| |
|---|
| Capturing screenshots |
| Enabling or changing administration settings |
| Opening and visiting any URL |
| Disabling Play Protect |
| Recording audio |
| Making phone calls |
| Stealing the contact list |
| Controlling the device via VNC |
| Sending, receiving and deleting SMS |
| Locking the device |
| Encrypting files on the device and external drives |
| Searching for files |
| Retrieving the GPS location |
| Capturing remote control commands from Twitter and Telegram |
| Pushing overlays |
| Reading the device ID |

Anubis targets mainly apps related to financial services, but also other shopping apps where monetization might be possible. The identified applications receive a slight update in the form of a fake overlay – again social engineering – to capture the user credentials (Feller, 2020). However, Anubis was not only distributed via phishing emails. It used/uses all available distribution channels as far this was possible and was indeed also available at the official Google App Store (Sun, 2019).

### 3.1.3 Mitigations and Limitations

Protection against malware could come in the form of shielding the mobile devices themselves against malicious code. This can be done by installing and keeping up-to-day anti-malware solutions on the device, having policies in place that prevent the download and installation of unsecured content, and being aware of the distribution channels is helpful to identify potential issues by the user. Unfortunately, in case apps from authorized app stores are indeed infected with malicious code, there is not much the end-user can do, except the standard mechanism that hopefully prevents the malware from causing harm as an antimalware solution on the mobile device. However, as the malware was already able to circumvent the protection mechanism from the authorized store there is a high likelihood that most malware solutions will not be able to identify it either. Overall, the best and easiest way to stop malware – as Anubis – is to prevent the installation in the first place as many malware types are often designed to specifically circumvent security mechanisms.

In the following sections, we will discuss phishing, communication, supply chain, and physical device breaches, which are often used as distribution channels for malware. One of the best ways to distribute malicious software as Anubis is to trick unaware end-users into taking certain actions. Therefore, we will first discuss social engineering, or more specifically phishing.

## 3.2 Phishing

Social Engineering is one of the most frequent attacks and everyone using the internet regularly has a high likelihood to have experienced it already. The most prevalent social engineering attack is phishing which affects anyone at any age and includes private and work scenarios. As a general rule in cybersecurity, humans tend to be the weakest link between the attacker and confidential information.

### 3.2.1 Phishing Types

Adversaries using phishing attacks disguise themselves as trustworthy parties and send fake emails or text messages, which might contain direct attachments (malware) or links to fake websites to deceive their targets (Fahl, 2019). The purpose of most phishing scams is to infect your device with malware, steal sensitive information like login credentials, or credit card numbers, obtain control of online accounts like email and social media or directly convince you – usually with a sense of urgency – to carry out a monetary transaction (Kaspersky, 2021a). The idea behind phishing is to exploit user interface weaknesses taking advantage of the fact that humans have difficulties with verifying URLs and dynamic content of rendered HTML documents (Fahl, 2019). Phishing usually serves as initial access to the victim's device or account, but often does not stop there. Since social engineering and phishing attacks depend largely on the generated trust, they can convey to the target it is even more successful to utilize the compromised accounts to launch a new more targeted attack on the contact list of the compromised account. Phishing attacks can result in identity theft or loss of money and can take different forms.

Phishing for the mass market – **spam phishing** – takes on a spam-like nature and bombards a large number of targets with emails or SMS, which usually contain links to forged websites, where the phishing attack continues. However, the email could also contain a direct download link. Attackers use forged websites to convince their victims they are interacting with the legitimate version. Usually, only a small percentage will fall for this trick, but given the huge volume, this will be enough to consider the attack a success. Spam phishing is easy and popular but tends to be easily identifiable as spam or malicious attempt (Kaspersky, 2021a).

More dangerous – especially for our CEO – are targeted **spear-phishing** attacks. Whaling usually takes on high-level targets from companies and government organizations, but can be applied to anyone the adversary deems a worthy target. **Whaling** attacks are planned rather long-term and start with an investigation/intelligence phase before preparing the attack itself. The amount of effort put into preparatory investigation and research can be significant before carrying out the actual attack. Attackers might gather details about their targets and organizations from publicly available information such as social media or earlier carried out data breaches. Due to the nature of whaling, which tailors the attack parameters specifically to the unique situation, every executive could fall for a well-prepared targeted phishing attack. To deceive our CEO into carrying out a specific pre-planned action the phishing adversary could compromise accounts of friends and family members of the CEO or colleagues working at the same company. We all tend to trust people we know and the unlikely nature of this attack would make it usually successful. What is dangerous about well-planned targeted phishing attacks is that they often do not stop. CEO fraud is a method of fraud in which the attacker pretends to be a senior executive and, for example, asks employees to transfer money to a specific account or to send confidential information. If the attacker uses e-mail as a means of communication, CEO fraud is a form of Business E-Mail Compromise (BEC) (Saud Al-Musib, Mohammad Al-Serhani, Humayun, & Jhanjhi, 2021). Usually faking the identity of an executive is called CEO fraud (Keyworth & Wall, 2016). This attack would be even more dangerous when the adversary had in fact access to the phone or email account of our CEO to convince the target to perform specific actions such as transferring funds or sending confidential information.

**Smishing** is the smartphone version of phishing and uses text messages such as SMS, whatsapp, and others instead of emails. However, the general logic remains the same. Two methods are used to steal the victim's data:

<u>Malware:</u> The Smishing URL link could directly convince you to download one of the malware types described in section 4.1. This would include all the techniques such as masquerading as a legitimate app, asking actively for permission, or confidential information.

<u>Malicious website:</u> Link to a forged website that mimics a legitimate one where you have to enter sensitive personal information (login credentials, credit card data, etc.)

A third way, which is especially relevant for our CEO scenario is direct action. In case it is possible to directly invoke the victim's trust, for example through an already compromised account due to an earlier attack, there is a possibility to convince the target into immediate action (e.g. transferring money, handing out login credentials, or any other information the attacker is after). For our CEO scenario, this is especially relevant as this "spear-Smishing" attack could easily result in immediate access to useful information.

### 3.2.2 Mitigations and Limitations

What is the best way for our CEO to prevent phishing attacks? What always should be in place for enterprises are potent anti-malware solutions, which can automatically quarantine suspicious files; network intrusion prevention mechanisms to block traffic at network boundaries; having restrictions in place for certain websites, which automatically block downloads/attachments or restrict browser extensions; utilization of anti-spoofing and email authentication instruments to filter messages based on validity checks of the sender domain and integrity of messages (MITRE, 2021). So much for the technical protection within enterprises, but this is nothing the end-user or employee can necessarily influence. Another way to prevent social engineering attacks is to apply "good judgment". If our CEO received an unusual request via the device or email of a close friend, family member, or colleague it might be best to give them a call to assure it was them who send the message. Social engineering is tricky and using common sense is one of the best mitigation tactics. Raising awareness of social engineering, education, and lifelong learning is key to staying afloat in this area. For example, not engaging with unknown emails, especially not clicking on URLs provided in emails or SMS. However, phishing attacks continuously expand and new variants evolve. Carefully planned spear-phishing attacks on high-profile targets can be almost impossible to avoid if the attacker chooses the right approach. There is always a chance that humans fall for a carefully planned social engineering attack. People are busy, which means we often simply do not have the "time or inclination" to think about security as we are preoccupied with other thoughts. After all, cyber-attacks are not a certainty, but more a likelihood that "could" happen if we are unlucky. From a business or academic standpoint, this attitude is referred to as cost-benefit analysis.

## 3.3 Communication

One of the major attack vectors on mobile devices is their wireless communication mechanisms. In this section, we will consider the usual suspects as Wi-Fi, Bluetooth, and Cellular networks. Communication channels requiring physical device access will be discussed in section 3.5.

### 3.3.1 Wi-Fi

Wi-Fi (802.11 networks) is one of the things in today's world we cannot live without. We can find Wi-Fi hotspots in all public places such as coffee shops, shopping malls, hotels, airports, and many others. Usually, these hotspots are free and do not require any passwords. This is very convenient for everyone, and this includes cybercriminals. Attackers gladly use these venues to intercept data from unaware victims to steal valuable information. Especially in today's world where everyone is carrying a smartphone, we might jump from one Wi-Fi hotspot to the next during a trip to a foreign city. Let's discuss some of the attacks our CEO could encounter.

Information exchange between two mobile devices over a network is done by sending packets back and forth. Capturing this traffic is easy as it is over the air and everyone connected to the same Wi-Fi has access to it. **Packet sniffing** can be done with open-source tools like Wireshark. If the traffic is in addition not encrypted it is readable. A more sophisticated option is setting up a **rogue Wi-Fi access point**. This is the case when an adversary either sets up a completely new Wi-Fi access point or compromises an already existing one. In case a victim connects to this malicious access point the attacker could eavesdrop or modify the network communication (MITRE, 2021). Setting up a rogue Wi-Fi access point is a very simple attack. Mobile devices automatically scan for and rank the strongest available Wi-Fi signal. Giving the rogue access point a trustworthy name will do the rest to convince many victims to connect without a second thought. Let's assume two parties want to initiate bilateral communication via a network in the hope to exchange information. When this communication is secretly intercepted by a cyber adversary, we refer to this as a **man-in-the-middle attack (MITM)**. The attacker redirects the traffic and can either eavesdrop on the conversation or impersonate one of the parties. (Weichbroth & Łysik, 2020). The major difference compared to packet sniffing is that the attack has also the option to alter the content of the packages. The adversary could change the email content or even insert malware.

Corporate executives as our CEO are especially vulnerable to all kinds of attacks on communication channels. Due to the potential high value of the information, they are carrying (e.g. potential M&A deals, financials, strategic initiatives, etc.) cybercriminals tend to put a lot more effort into successfully carrying out an attack. Therefore, adversaries tend to focus specifically on venues that a regularly frequented by business executives. A very good example is the "Darkhotel" spy campaign in luxury Asian hotels (Drozhzhin, 2014). The attackers compromised several high-class hotels. Once an executive checked into the hotel and connects to the room-Wi-Fi the attack starts. Hotels tend to be very similar when it comes to login requirements, which are the room number and the surname of the guest. Upon login, the attacker initiated a software update. Instead of updating Google, Facebook, or another well-known application the attackers installed a backdoor on the mobile device of the executive. Once the malware was installed the attacker can collect data, passwords, and other login credentials. Darkhotel was not a one-time attack, but an ongoing cyberespionage campaign. It is important to mention that this was an organized campaign that utilized many zero-day vulnerabilities. The attackers also used different malware types and further phishing attacks once the initial backdoor was installed. Also, the time horizon of the attack is quite astonishing. "The #Darkhotel campaign appears to have been active for seven years" (Drozhzhin, 2014).

Our CEO will be pleased to hear that the threat through public Wi-Fi attacks can be mitigated. The easiest way is not to use public Wi-Fi hotspots at all and simply use the cellular network instead. The prerequisite would be sufficient mobile data to be independent of random Wi-Fi access points. Another option is to use a virtual private network (VPN). This is something that should be a permanent solution and not only because employees or executives want to use public Wi-Fi. Home networks are usually not subject to strong security mechanisms either. A company could with mobile devices management (MDM) solutions control the Wi-Fi access points that employees are allowed to connect to. Security considerations are important, however, it is not a good idea to sabotage useability and end-user satisfaction with unnecessary and especially "annoying" restrictions. Remote work will significantly increase during the coming years and the "Digital Nomad" lifestyle will with a high likelihood establish itself as the new normal. In this world where everyone is changing locations frequently – maybe even several times per day – there should be a reliable solution for mobile devices in place at all times. And this solution needs to be user-friendly. Next to the VPN tunnel should be an up-to-date firewall and antivirus solution in place. In addition to the technical solutions such as VPN, firewall, and anti-malware solutions, employees which carry mobile devices should receive security training that raises the general awareness level regarding potential cyber-attacks. Treating all kinds of unsecured public Wi-Fi access points with a little suspicion might already help. Limitations are of course that a rogue access point could even include access to the corporate VPN. This could be achieved with the help of an insider (e.g. an employee of the company).

### 3.3.2   Bluetooth and Cellular

Bluetooth is a network technology for short-range radio frequency that enables mobile devices to communicate with one another. Bluetooth-enabled devices are mostly used as personal devices where sensitive data are stored. Exposing these sensitive data can be an issue for a person's safety. Similar to Wi-Fi, devices communicating via Bluetooth are susceptible to wireless networking threats. The usual suspects are denial of service (DoS) attacks, eavesdropping, man-in-the-middle (MITM) attacks, message modification, and resource misappropriation. In addition, there are more specific attacks specifically related to the Bluetooth technology itself (Padgette & Padgette, 2017). Examples of Bluetooth vulnerabilities/attacks are Bluejacking, Bluesnarfing,

Bluebug, Cracking PIN, Blueprinting, BlueBorne, Bleeding Bit, KNOB, and SweynTooth (Hargreaves, 2021).

To secure Bluetooth-enabled devices organizations are advised to choose the strongest security mode available – there are different kinds based on the specific Bluetooth version (Padgette & Padgette, 2017). Cope et al. (2017) analyzed Bluetooth security vulnerabilities and came up with 17 recommendations to improve the security of those devices. Examples are disabling Bluetooth when the device is not in use; using adequate encryption key strength to prevent brute force attacks; setting the device to undiscoverable by default; utilization of encryption to prevent eavesdropping. Overall, "*As with any wireless technology, there is no way to prevent all attacks and guarantee security* (Cope et al., 2017)." Nevertheless, user awareness training as mentioned already several times can help to mitigate attacks that are mainly based on fooling humans.

Another way next to Wi-Fi or Bluetooth is to use the **cellular network** for attacks. The major threat here is the exploitation of vulnerabilities in the global mobile phone system to snoop on calls, and texts, or snatch the device location. In case the attacker sends in SMS via radio interfaces this would count as a cellular attack but is essentially Smishing, which might contain links to compromised websites or direct downloads of malware. Also interesting is an SS7 Attack, which might either be used to track the location of the device or to redirect calls or SMS to a number controlled by the adversary. Once this happened, the adversary can act as a man-in-the-middle to intercept and alter the communication. This technique could be used to steal authentication codes for multi-factor authentication. Similar to the Wi-Fi version an adversary could set up a rogue cellular base station for eavesdropping or manipulating cellular device communication (MITRE, 2021).

Mitigations are here similar, which largely consist of making sure to encrypt communication. In addition, network carriers could use firewalls, and intrusion detection/prevention systems to fight SS7 exploitation (MITRE, 2021).

### 3.3.3 Denial of Service

Alternatively, especially when the purpose is not to access the device, attackers could simply jam radio signals like Wi-Fi, Bluetooth, and Cellular. (Distributed) Denial of Services (DOS) attacks are a general way to flood the system of the target with an overload of bogus traffic with the purpose to slow down or crash the system. DDoS attacks use so-called zombies, which are infected devices that form a botnet. Using a large number of zombies will significantly increase the strength of the denial-of-service attack. Also, attackers could try to downgrade the communication to less secure protocols. Jamming frequencies used by newer protocols such as LTE might force mobile devices to utilize older protocols for communication (MITRE, 2021). This will help the attacker to carry out the above-mentioned attacks such as packet sniffing, rogue access point, and man-in-the-middle. Jamming or Denial of Service attacks cannot be easily mitigated.

## 3.4 Supply Chain

A supply chain attack aims to harm a business by targeting weak elements in the supply chain. A supply chain attack can occur in any industry. Cybercriminals often manipulate the manufacturing process of a product by installing a rootkit or hardware-based spying components. In this section, we will differentiate between physical and app store ecosystem-related supply chain attacks.

### 3.4.1 Physical Supply Chain

Cybercriminals can manipulate hardware components in mobile devices along the supply chain before the final shipment to the customer to achieve a data or system compromise. This is often achieved by inserting a backdoor in the system that later grants the attacker unauthorized access. A backdoor is an alternative method of access to programs or hardware systems that bypasses the usual security mechanisms (MITRE, 2021).

Physical supply chain compromises cannot be addressed by end-user whether for private or business purposes. The best option for suppliers is tight control of their supply network to prevent possible harm from cybercriminals. One option our CEO has that might reduce the likelihood of receiving a compromised device is to focus on high-end phones from well-known and trusted suppliers due to more sophisticated security controls. Another one is to perform a physical inspection of hardware to look for potential tampering upon delivery of the new device.

### 3.4.2 App Store Ecosystem

As already touched upon in section 3.1, one of the core distribution channels for malicious software or code to the device of an end-user is through compromising the app store supply chain. Software companies tend to ship new applications and updates thereof daily via the internet and cloud. Therefore, adversaries try to place the malicious application in an authorized app store, enabling the application to be installed on the victims' devices (MITRE, 2021).

A very good example of a supply chain-delivered malware is SimBad. This adware campaign was discovered by Check Point Software Technologies in 2019. It was distributed over the official authorized Google Play Store, was spread across 206 applications and affected 150 million users. The name SimBad stems from the fact that most infected apps were simulator games and transformed the use of the mobile device into an unbearable experience by displaying countless ads outside of the application, with no visible way to uninstall them (Root & Polkovnichenko, 2019). SimBad had the following malicious behavior (Clayton, 2019):

**Figure 4 – SimBad – Malicious Behaviour**

| | |
|---|---|
| 1 | Showing ads outside of the application, for example when the user unlocks their phone or uses other apps. |
| 2 | Constantly opening Google Play or 9Apps Store and redirecting to another particular application, so the developer can profit from additional installations. |
| 3 | Hiding its icon from the launcher in order to prevent uninstallation. |
| 4 | Opening a web browser with links provided by the app developer. |
| 5 | Downloading APK files and asking the user to install it. |
| 6 | Searching a word provided by the app in Google Play. |

Modern mobile apps often utilize several third-party libraries and open-source components, which is important for DevOps teams to speed up the creation and deployment of applications. SimBad found its way into the supply chain via a Software Development Kit (SDK) for an ad-related SDK. Developers globally were somehow convinced to use this SDK, were not aware of its content, and uploaded the final and malware-infected application onto the Google Play Store.

What are recommended mitigations? First and foremost, users should be encouraged to only install apps from authorized app stores and keep the natural protection mechanisms active that are provided by the vendors. For example, there is a setting on Android called "unknown source", which needs to be active for users to install 3rd party apps. Hence, users should be discouraged to enable that setting. In addition, to shield against malware that is delivered via 3rd party app stores the installation of unsigned applications should not be allowed. This could be enforced through corporate policies and a mobile device management (MDM) system that ensures that only pre-approved applications can be installed on work devices (MITRE, 2021). In case the company operates under a Bring Your Own Device (BYOD) policy, the MDM will only be able to control the work-space part of the device (Boeckl et al., 2021). See section 4.1 for more details.

Even though there is no guarantee to avoid malicious apps it is less likely that the official store contains malicious repackaged apps. Google, for example, is using Google Play Protect to screen for potential dangers. Google Play Protect is a protection service for all android devices which continuously scans Google Play Store and user devices for malware (Google, 2021b). According to Google (Google, 2021b): *"All Android apps undergo rigorous security testing before appearing in Google Play. Google Play Protect scans more than 100 billion apps daily to make sure that everything remains spot on. That way, no matter where you download an app from, you know it's been checked by Google Play Protect."* It is a useful protection mechanism and helps with increasing the security levels of the smartphone, but as shown in the examples in section 3.1 (Anubis) and 3.4 (SimBad) it is often not enough. The first step during the Anubis installation is disabling Google Play Protect, which renders it immediately worthless. Here again, since the end-user specifically had to give access to Anubis to disable Google Play Protect, being slightly suspicious helps to prevent many social engineering attacks.

## 3.5 Physical and Authentication

### 3.5.1 Physical Device breaches

It was already mentioned that the human end-user tends to be the most serious security risk when it comes to installing malware on their devices due to a lack of awareness of how social engineering works and also the general inability of humans to spot errors on HTML-rendered user interfaces. This is equally true for this part. The greatest threat to mobile devices and the associated loss of data is usually the owner of the mobile device. It is significantly more like for users to simply lose their phone compared to it being stolen (Weichbroth & Łysik, 2020). Another threat is physical storage devices like USBs and external hard drives, which can be connected to a smartphone. The same thing is true for compromised charging stations in public areas. Attackers can use these physical connections to install malware on the victim's device. As described in section 3.1 Malware, most types for this purpose are quite persistent and evade detection. They tend to be self-replicating and spread to other devices in the network secretly (worm) (Abijah Roseline & Geetha, 2021).

Mobile device owners should avoid using public charging stations or computers and only use their own charger or computer acquired from a trustworthy source. Also, it is vital not to click on any permission requests. For example, confirming that the attached computer or another device can be trusted. Mitigation techniques for the "stolen or lost" threats are easy to describe, but difficult to enforce: simply do not lose your device or get it stolen. Especially in public areas such as coffee shops, devices should not be left unattended, which is part of the mitigation technique "common sense". The second major mitigation technique against stolen or lost phones are strong authentication mechanisms, which are discussed next.

### 3.5.2 Passwords and Biometrics

The most common mechanism for smartphone authentication are passwords, PINs, patterns, and biometrics (Fahl, 2019). The most secure option is a long and complex password but forcing the end-user to enter a long and complex password every time the screen is turned off is pure torture. Most mobile device users choose easier methods as short PINs (4-digits minimum) or a pattern (3x3 grid) to unlock the screen, which is significantly less secure. In addition to being vulnerable to shoulder surfing, unluck patterns have quite low entropy as users tend to choose rather similar versions (Fahl, 2019). Another option is to use our fingerprint or facial recognition. Biometric features do not replace the traditional password but offer us a more convenient way of accessing our mobile devices. The main advantage of biometrics is that it allows us to choose an extra-strong password. Facial recognition allows attackers often to the use a simple picture to unlock the device. And if the 3D effect is required a 3D printed version might do the trick. The strongest combination is a complex password in combination with the fingerprint biometric feature for convenience (Samsung for Business, 2021). Physical attacks include smudge attacks and shoulder surfing. Smudge attacks are based on the oily smudges that the victim's fingers have left behind. Here it is already obvious, that a complex password with overlapping characters will be significantly more difficult to read from this oily surface than a pattern or short PIN. Shoulder surfing usually uses dedicated equipment like cameras or telescopes and occurs often in public places similar to Wi-Fi interception attacks. Shoulder surfing might also capture login information to websites and other sensitive information from the screen.

An adversary with physical access to a mobile device is in the position to attempt to bypass the device's lock screen. The attacker could try to unlock the device through pure guessing or brute force password attacks. Another option would be biometric spoofing. The attacker could create a 3D wax face (Jia, Hu, Li, & Xu, 2021) or simply use a picture for facial recognition. For the fingerprint, a fake silicon finger could do the trick.

Companies have different options to decrease the likelihood of a physical device breach. One option is to increase the password complexity by requiring a certain password length, special characters, etc. is one option to make a simple brute force attack more difficult. Delays between the different password attempts are also standard. In addition, it is recommended to automatically wipe all the data if a wrong password is entered incorrectly too many times. Both of these policies would reduce the threat of brute force attacks, simple password guessing, and shoulder surfing (MITRE, 2021). To avoid biometric spoofing mobile phone vendors require the user after a certain amount of idle time to enter the password instead of only using biometrics. This is also true after every device restart. This helps to prevent biometric spoofing in case of a loss or a stolen phone. "Securing facial recognition: the new spoofs and solutions" is an article from Hassani and Malik (2021) that goes here in more detail.

# 4 Security Design and Future Research

## 4.1 Modern BYOD MS Design

In this part, we will have a look at the mechanisms required to achieve appropriate security and privacy levels for workplace devices, which is also recommended for our CEO. Due to the popularity of BYOD policies, mobile devices in corporate environments have become an omnipresent reality. Carrying two or more devices (one for private use and one for work) is not convenient and the movement towards a BYOD mindset should be looked upon in favor. Accepting this new reality will help to speed up the implementation of new security measures in most corporations across the globe. To support the security and privacy goals of BYOD devices the following mechanisms are recommended (Boeckl et al., 2021): Trusted Execution Environment, Enterprise Mobility Management (EMM), which is a system that can be used to provision policies to mobile devices to control aspects of their allowed behavior, Virtual Private Network (VPN), Mobile Application Vetting Service, and Mobile Threat Defense, which may be able to spot and prevent the installations from sources outside of the official app stores. Those mechanisms will achieve the following benefits (Boeckl et al., 2021):

1. Separation of organizational and personal information by creating a logical separation between personal and workspace is achieved;

2. Encrypting data in transit with the help of a VPN or similar solutions (having the right certificates is a prerequisite for accessing the companies' resources) to reduce the risk of nonsecure networks, which put data at risk of interception;

3. Devices can identify vulnerable applications and also react by removing the employer's work profile, while simultaneously keeping the personal information untouched;

4. Malware detection is in place to spot malicious software, especially from 3rd party stores (see section 3.1);

5. Trusted access due to security certificates and maybe other user credentials to ensure two-factor authentication; and finally, assuring end-user privacy by restricting information collection to avoid revealing sensitive personal data, device location, etc.

The core feature, which is the logical separation of personal space and workspace on the same device has several advantages. Modern devices are Android and iOS are already capable of the following (Boeckl et al., 2021; Google, 2021a): (1) data flow restriction between enterprise and personal applications, (2) restriction of application installation from unknown sources, (3) selective wiping to remove enterprise data and preserve personal data, (4) device passcode requirement enforcement, (5) application configuration control, (5) identity and certificate authority certificate support.

It is vital to understand that risk assessments are always a likelihood estimate of a negative outcome. This futuristic scenario might never materialize, which means that most actions in cybersecurity are a trade-off and subject to cost-benefit considerations. This reality might result in systems with weak security mechanisms never experiencing any attack and cyber fortresses being breached by an over-motivated and capable hacker, who utilized the latest zero-day exploit.

However, given that today's mobile devices already possess the capabilities described above I would be recommended that our CEO or the company for that matter assures us to implement those security measures. We have seen in section 3 'Attacks and Mitigations' that cybercriminals are creative and keep expanding the pool of attacks consistently. Especially in today's world, which is becoming more digital by the minute will the importance of cybersecurity only increase. It has already become one of the most important – and unfortunately lacking – skillsets in the industry, but recurring global attacks (e.g. the Kaseya supply chain ransomware attack), which cause significant damage to enterprises increasingly help to convince decision-makers to take action (Allgeier, 2021).

## 4.2   Securing the CEO's mobile device

Let us summarize the main ideas regarding device protection. Overall, it depends on the concrete situation. Larger companies will take care of their own devices and will often utilize mobile device management solutions that take care of all the end-user security concerns. Also, it might be that only company approve applications are free to download. For start-ups and smaller companies, the situation is different. Here, the employee could be completely responsible – similar to the private device – to take care of the smartphone entirely, including all matters of security. Here are some practical recommendations for securing our CEO's mobile devices: Use strong passwords/biometrics for authentication; Avoid public Wi-Fi / unencrypted connections; Utilize a Virtual Private Network (VPN); Encrypt your device or move important data to the cloud; Install an antivirus/antimalware application; Keep your device up-to-date.

Keeping the above-mentioned tips in mind the likelihood to fall for a random spam-like attack will be significantly reduced. More targeted attacks from powerful adversaries, however, will always be dangerous, but the likelihood to be the victim is probably rather low.

### 4.3   Future Topics in Mobile Security

The most significant threats for mobile devices keep changing every year (GSMA, 2021), and marching into the next decade new technological innovation and product introductions will accelerate, which will further expand the attack surface (Richardson, 2020). The major focus of this paper was on smartphone security with a specific focus on a hypothetical CEO scenario. However, mobile security goes beyond smartphones. It includes – subject to the concrete definition – a while basked of devices, which will gradually be present in almost any device we can imagine. Maybe in the not-too-distant future, we will become mobile devices ourselves - microchips could be implemented below our skin. We see, it becomes vital to have absolute security when adversaries could potentially even access humans themselves and gain access to their brain-computer interfaces, which are potentially steering other remote devices carrying out vital tasks. The following fields might be interesting to explore further:

**Zero Trust** is one interesting research direction in mobile security (Haddon, 2021). The workforce will increasingly be flexible and remote and wants to be secure everywhere (Sheridan, 2021). Also, customer expectations are rising consistently in our digital age. This means, that the more seamlessly security can be integrated into our devices without creating additional effort for the end-user the better. **Mobile Cloud and Blockchain**, or in general securing distributed computing environments. Mobile phones as the frontend are only the interface to our data, and due to "X as a Service" solutions soon to almost everything (Li et al., 2021). Smartphones and the gaming industry might for example significantly benefit from "Computing Power as a Service" offerings. **Artificial Intelligence** has found its way already in every industry. Nevertheless, AI has the potential to be used in both an exploitative and positive sense. AI as general-purpose technology can be utilized by adversaries and security professionals alike to achieve their purpose. AI can help to generate new attacks, but also enhance existing security mechanisms to shield individuals, companies, and governments from cybercriminals (Sarker, Furhad, & Nowrozy, 2021). Interesting examples are Dark Trace and Siemens AI-Driven Cyber Security. **Quantum Computing** is still in its infancy, but Quantum-Safe Cryptography will become increasingly important as the capabilities and the availability of those systems increase (Mitchell, 2020).

## 5   Summary and Conclusion

Mobile devices have become omnipresent and Covid 19 catapulted us into a new age, which is characterized by a fully remote and therefore flexible workforce. Our digital age and the increasing importance of mobile devices have also attracted cyber criminals that are after valuable information. Malware – which is malicious software installed on our devices – is one of the core concepts present in cybersecurity and plays a huge part in many attacks. There are several types of malware including trojans, viruses, and worms, which are often self-replicating and actively try to hide from security mechanisms to stay permanently on the device once installed. Malware can be distributed via different channels, which often goes hand-in-hand with the other attacks discussed. Malware can be distributed via phishing attacks. Those can have a spam-like nature catering to the mass market, or be more targeted to specific individuals. Especially relevant for high-profile individuals as C-level executives are the more focused variants of spear-phishing or whaling. Another way to gain access to victims' devices and information is via network attacks that utilize our communication channels such as Wifi, Bluetooth, or cellular networks. Packages within those networks can eavesdrop, or the communication can be completely intercepted via a man-in-the-middle attack, which allows the attacker not only to listen but also to alter the content of messages. Due to our end-to-end physical as well as digital supply chains, it is possible to place malware into devices and applications before they reach the end-user. Authorized app stores such as google play have several security measures in place, but this does not prevent adversaries from successfully placing malware-infected applications onto the store. Last but not least, it turns out that people are one of the weakest links when it comes to security as the likelihood that a device is stolen or compromised is significantly less than that of losing a smartphone. Therefore, it is highly recommended to take authentication mechanisms such as passwords and biometrics seriously and avoid utilizing patterns and four-digit pins to unlock our devices. Moreover, once breached, mobile devices can serve as an entrance point for the enterprise backend and more serious security compromises thereafter. Not everything is bleak though.

The BYOD movement has increased the usage of mobile devices for work globally and triggered new security features for smartphones. The logical separation of personal space and work assures the security of corporate information, while at the same time hiding private information like personal data and device location. In case of the identification of malicious activity, the workspace can be remote wiped while keeping the personal space untouched. Overall, a zero-trust security approach is a way to go in our ever more complex, global, and digital world economy. People are already used to perfect flexibility, which means not thinking about security during already busy schedules will be favored. Therefore, fully managed end-points using a zero-trust policy, separated private and mobile spaces, VPN connections, etc. as a standard. Other topics such as IoT, AI, Cloud, and Blockchain will keep integrating and further seamlessly combining every part of the digital economy.

The only constant in our digital world is "change". Innovations, breakthroughs, and human creativity will continuously create new attack vectors, which increases the overall attack surface.

Cyber Security will remain a constant struggle against highly intelligent adversaries, which are consistently striving to improve upon the status quo. And we – and our CEO – should strive to be prepared.

Mobile Security: Attacks and Mitigations                                                                 24Hargreaves, C. (2021). *Mobile Communications Security (Lecture Slides). Mobile System Security (MSS)*. University of Oxford.

Hassani, A., & Malik, H. (2021). Securing facial recognition: the new spoofs and solutions. *Biometric Technology Today*, *2021*(5), 5–9. https://doi.org/10.1016/S0969-4765(21)00059-X

Jia, S., Hu, C., Li, X., & Xu, Z. (2021). Face spoofing detection under super-realistic 3D wax face attacks. *Pattern Recognition Letters*, *145*, 103–109. https://doi.org/10.1016/j.patrec.2021.01.021

Kaspersky. (2021a). All About Phishing Scams & Prevention: What You Need to Know. Retrieved from https://www.kaspersky.com/resource-center/preemptive-safety/phishing-prevention-tips

Kaspersky. (2021b). Types of Malware. Retrieved from https://www.kaspersky.com/resource-center/threats/malware-classifications

Keyworth, M., & Wall, M. (2016). The "bogus boss" email scam costing firms millions. Retrieved from https://www.bbc.com/news/business-35250678

Lee, W. (2019). Malware and Attack Technologies. *The Cyber Security Body of Knowledge*.

Li, G., Ren, X., Wu, J., Ji, W., Yu, H., Cao, J., & Wang, R. (2021). Blockchain-based mobile edge computing system. *Information Sciences*, *561*, 70–80. https://doi.org/10.1016/j.ins.2021.01.050

Mitchell, C. J. (2020). The impact of quantum computing on real-world security: A 5G case study. *Computers and Security*, *93*, 1–11. https://doi.org/10.1016/j.cose.2020.101825

MITRE. (2021). MITRE ATT&CK®. Retrieved from https://attack.mitre.org/

Padgette, J., & Padgette, J. (2017). Guide to Bluetooth Security. *NIST Special Publication 800-121 Revision 2*, 1–67.

Richardson, J. (2020). Is there a silver bullet to stop cybercrime? *Computer Fraud and Security*, *2020*(5), 6–8. https://doi.org/10.1016/S1361-3723(20)30050-6

Root, E., & Polkovnichenko, A. (2019). SimBad: A Rogue Adware Campaign On Google Play. Retrieved from https://research.checkpoint.com/2019/simbad-a-rogue-adware-campaign-on-google-play/

Samsung for Business. (2021). Which biometric authentication method is most secure? Retrieved from https://insights.samsung.com/2021/05/25/which-biometric-authentication-method-is-most-secure-3/

Sarker, I. H., Furhad, M. H., & Nowrozy, R. (2021). AI-Driven Cybersecurity: An Overview, Security Intelligence Modeling and Research Directions. *SN Computer Science*, *2*(3), 1–18. https://doi.org/10.1007/s42979-021-00557-0

Saud Al-Musib, N., Mohammad Al-Serhani, F., Humayun, M., & Jhanjhi, N. Z. (2021). Business email compromise (BEC) attacks. *Materials Today: Proceedings*, (xxxx). https://doi.org/10.1016/j.matpr.2021.03.647

Sheridan, O. (2021). The state of zero trust in the age of fluid working. *Network Security*, *2021*(2), 15–17. https://doi.org/10.1016/S1353-4858(21)00019-2